\documentstyle[aps,prd]{revtex}

\newcommand{\HH}{{\cal H}}

\newcommand{\de}{\delta}

\newcommand{\ep}{\epsilon}

\newcommand{\ka}{\kappa}

\newcommand{\be}{\begin{equation}}
\newcommand{\ee}{\end{equation}}

\newcommand{\bea}{\begin{eqnarray}}
\newcommand{\eea}{\end{eqnarray}}
\newcommand{\bean}{\begin{eqnarray*}}
\newcommand{\eean}{\end{eqnarray*}}

\def\spose#1{\hbox to 0pt{#1\hss}}
\def\ltapprox{\mathrel{\spose{\lower 3pt\hbox{$\mathchar"218$}}
\raise 2.0pt\hbox{$\mathchar"13C$}}}
\def\gtapprox{\mathrel{\spose{\lower 3pt\hbox{$\mathchar"218$}}
\raise 2.0pt\hbox{$\mathchar"13E$}}}

\begin{document}
\draft
\preprint{\ 
\begin{tabular}{rr}
& 
\end{tabular}
} 
\twocolumn[\hsize\textwidth\columnwidth\hsize\csname@twocolumnfalse\endcsname 
\title{Clarifying perturbations in the ekpyrotic universe\\
  A web-note}
\author{Ruth Durrer}
\address{ Institute for Advanced Study, Einstein Drive, Princeton, NJ 08540, 
	USA\\  and \\  
D\'epartement de Physique Th\'eorique, Universit\'e de Gen\`eve,
24 quai E.\ Ansermet, CH-1211 Gen\`eve 4, Switzerland}
\maketitle

\begin{abstract}
In this note I try to clarify the problem of perturbations in the 
ekpyrotic universe. I write down the most general matching conditions and
specify the choices taken by the two debating sides. I also bring up the
problem of surface stresses which always have to be present when a transition
from a collapsing to an expanding phase is made.
\end{abstract}
\date{\today}
] 
\renewcommand{\thefootnote}{\arabic{footnote}} \setcounter{footnote}{0}
\section{Introduction}
Lately, the idea that a hot big bang universe might emerge from the collision 
of two BPS 3-branes in $4+1$ dimensions has been put 
forward~\cite{ekpy1,nathi,ekpy3}. For an observer on a brane, the 
universe is first contracting 
and then goes through a big bang into a hot, thermal expanding 
phase. During the contracting phase, the matter content is dominated by a 
scalar field, the amplitude of which is physically related to the distance 
to the second brane. It has been argued that, under certain conditions for 
the scalar field potential, this scenario can lead to a scale invariant 
spectrum of scalar perturbations. Several papers have objected to this 
result~\cite{david1,robi,david2}, mainly claiming 
that the scale invariant growing mode before the collision has to be matched 
to the decaying mode after the collision. The decaying mode before the 
collision, which then is matched to the growing mode after the collision is 
blue and will therefore not lead to a scale invariant spectrum of scalar 
fluctuations. 

These objections have come from the usual matching conditions
after inflation~\cite{nathalie}, where the extrinsic curvature on surfaces
of constant energy density. However, the matching
surfaces might be chosen differently, and I will show that by choosing 
another matching surface so that the energy perturbations in comoving 
gauge are continuous during the transition, the growing 
mode of the contraction phase is in general inherited by the 'growing' 
mode of the expansion phase.

Furthermore, in the ekpyrotic universe the matching
has to be done from a collapsing phase into an expanding phase. Here,
the extrinsic curvature of the background changes sign. Having a jump it 
the background extrinsic curvature, it does not seem reasonable 
to require that its perturbation be continuous. Clearly, within the four 
dimensional picture, for the  extrinsic curvature
 to jump, we have to require a non-vanishing surface stress 
density. The perturbations of which will enter the matching conditions.

In this note, I discuss the matching conditions applied in~\cite{ekpy3} and
\cite{david1,robi,david2}.  It appears to me that the matching conditions 
of~\cite{ekpy3} are more natural that those applied in 
\cite{david1,robi,david2}, where the surfaces of constant energy density
 are matched and
the perturbations of the surface stress density is set to zero.
The truly correct matching conditions can only be worked out in the five 
dimensional picture which has to determine this surface stress density   
and its fluctuations.

Here I shall remain fully in within the four dimensional framework and just 
discuss the generic matching conditions which I then exemplify in the two 
cases studied in the literature. In one of them, where $\delta \rho/\rho'$
in longitudinal gauge is continuous , the scale invariant growing mode of 
the collapsing phase is matched entirely to the decaying mode in the 
expanding phase.
The second case, where the energy perturbation $\delta \rho$ in comoving 
gauge is continuous, leads to a growing mode in the expanding phase 
which is inherited from the growing mode in the collapsing phase.

Since this is a web-note, mainly addressed to those
who have studied the ekpyrotic scenario and have been puzzled about the 
ongoing perturbation debate, we shall not repeat the basics of the scenario,
but immediately jump into 'medias res'.

\section{Matching perturbations from a collapsing to an expanding universe}
Let us first note some generic results about matching of perturbations, which
are especially subtle when matching from a contracting to an expanding phase.

The extrinsic curvature in a Friedman universe is
\be
  K^i_j = -\left({\dot a\over a^2}\right)\de^i_j = - 
  {{\cal H}\over a}\delta^i_j~,
\ee
where $a$ denotes the scale factor and an over-dot is the derivative w.r.t. 
conformal time $\eta$ and ${\cal H}=\dot a/a$.  $\dot a$ changes sign in 
the transition from a contracting to an expanding phase. Hence the  
 extrinsic curvature is 
discontinuous in the four dimensional picture, if we simply 'glue' the 
contracting phase at very small scale factor (high curvature) to the
expanding phase with the same scale factor, but opposite sign
for $\dot a$ and conformal time $\eta$. Clearly, this is the simplest 
way of connecting a contracting phase to an expending phase, but it is 
relatively close to the approach motivated from the 5 dimensional picture, 
where the singularity at $a=0$ becomes a narrow 'throat' \cite{nathi}. Here
we replace this throat by a stiff 'collar' (see also \cite{ekpy3}). 

If the extrinsic curvature is discontinuous, the Israel junction 
conditions require~\cite{isi}
\begin{equation}
\left[K^i_j\right]_\pm=\kappa^2 S^i_j,
\end{equation}
where $S^i_j$ is a surface stress--energy tensor, and 
$$[h]_{\pm} \equiv   \lim_{\ep\searrow 0}[h(\eta_1 +\ep) -h(\eta_1-\ep)]
  = h_+ - h_-$$
for an arbitrary function $h(\eta)$. The time $\eta_1$ is the time of 
matching.
Let us consider an arbitrary hyper-surface, linearly perturbed 
from $\eta=$const. and defined by $f=f_0(\eta) +\de f =$ const., where $f$ is 
an arbitrary (perturbed) function, e.g. the energy density.
We use longitudinal gauge,
\be
 ds^2 = a^2(\eta)[-(1+ 2\Psi)d\eta^2 + (1+ 2\Phi)\de_{ij}dx^idx^j]
\ee
(we ignore a possible spatial 3-curvature which is unimportant at early time).
The perturbation variables $\Psi$ and $\Phi$ are the Bardeen potentials. We 
assume that there are no anisotropic stresses, so that $\Phi=-\Psi$. We want to
determine the extrinsic curvature in the coordinate system $(\tilde\eta, 
\tilde x^i)$ where the surfaces $\{ \tilde\eta=$ const. $\}$ are parallel 
to $\{ f=$ const. $\}$. The perturbed conformal time is given by 
$\tilde\eta =\eta +\de f/\dot f$
The normal vector to $f=$ const. in this coordinate system is \cite{nathalie}
\begin{equation}
\tilde n_0=-a(1+\Psi +{\cal H}\de f/\dot f), \qquad 
   \tilde n_i=-a\partial_i\delta f/\dot f~.
\end{equation}
The induced metric and extrinsic curvature in the coordinate system 
$(\tilde\eta, \tilde x^i)$are given by
by
\begin{equation}
\tilde q_{\mu\nu}\equiv \tilde g_{\mu\nu}+\tilde n_\mu \tilde n_\nu,\qquad
\tilde K_{\mu\nu}\equiv\tilde \nabla_{\mu}\tilde n_\nu
\end{equation}
where $\tilde g_{\mu\nu}$ is the metric and $\tilde \nabla$ the covariant 
derivative. The continuity of the induced metric 
implies (for the background) that
\begin{equation}\label{m0}
\left[ a(\eta)\right]_\pm=0~.
\end{equation}
The background  extrinsic curvature is
\begin{equation}
K^i_j=-\frac{{\cal H}}{a}\delta^i_j~.
\end{equation}
Since ${\cal H}$ changes sign at the transition, $K^i_j$ is not continuous 
and we have
\be
\left[K^i_j\right]_\pm= -2\frac{{\cal H}_+}{a}\delta^i_j = 
\kappa^2 p_s\de^i_j, \label{backjump}
\ee
where $p_s$ is a negative surface pressure. Within the four dimensional 
picture we have no explanation for this surface pressure, which has to be 
present in order for the extrinsic curvature to jump, if we require 
Einstein's 
equation to hold during the transition. (If they don't hold, we have 
no means to find the matching conditions in a four dimensional analysis.) 
Eq.~(\ref{backjump}) is a possibility to 'escape' the violation of the weak
energy condition, $\rho+p < 0$, which is needed for a smooth transition from 
collapse to expansion. This has been one of the objection to the ekpyrotic
scenario~\cite{linde}.

At the perturbed level, the continuity of the induced
metric leads to \cite{nathalie}
\begin{equation}\label{m1}
\left[\Psi+{\cal H}\frac{\delta f}{\dot f}\right]_\pm=0~.
\end{equation}
 The extrinsic curvature is given by~\cite{nathalie}
\begin{equation}\label{m2}
\delta K^i_j=\frac{1}{a}\left[\dot\Psi+{\cal H}\Psi
+(\dot{\cal H}-{\cal H}^2)\frac{\delta f}{\dot f}
\right]\delta^i_j. +(\de f/\dot f)^{,i}_{,j}
\end{equation}
For simplicity, we assume that also the surface stress tensor has no 
anisotropic stresses, which implies 
\be[(\de f/\dot f)]_\pm=0 ~.  \label{mani} \ee  Then
\begin{equation}
\delta K^i_j\equiv(\delta K)\delta^i_j
\end{equation}
so that the matching conditions for the perturbations become (\ref{mani})
and 
\begin{equation}\label{m3}
\left[\delta K\right]_\pm=\kappa^2\delta p_{s}~, 
\end{equation}
where $\de p_{s}$ is the perturbation of the surface pressure.

Let us first consider the constant energy hyper-surfaces in longitudinal 
gauge, which are usually used for matching after inflation, and on which all 
attention of ~\cite{david1,robi,david2} has been concentrated. In this case
 $f=\rho$ and $\de f=\de \rho$ in longitudinal gauge. The perturbed 
Einstein equations give
(see e.g.~\cite{du94} Eqs. (2.45), (2.46) and use $\de\rho= \rho D_s$ in 
 longitudinal gauge)
\be
{\de\rho\over \rho} = {-2\over {\cal H}^2}\left[(3k^2+{\cal H}^2)\Psi 
	+{\cal H}\dot\Psi \right] \simeq  - 2\left[\Psi +
	{\cal H}^{-1}\dot\Psi \right] 
\ee
on super horizon scales.
With $\dot\rho =  2{\dot\HH-\HH^2\over \HH}\rho $
this yields
\be
 \de \rho/\dot\rho \simeq {-1\over \dot\HH-\HH^2}(\HH\Psi 
  +\dot\Psi)~. \label{mani2}
\ee
Eq.~ (\ref{m1}) then leads to
\begin{equation}\label{me1}
\left[ \Psi-  {\HH\over \dot\HH-\HH^2}[\HH \Psi + \dot\Psi]\right]_\pm
= [\zeta]_\pm = 0~,
\end{equation}
where $\zeta$ is the curvature perturbation introduced by Bardeen.
Note that in general $\Psi$ will not be continuous at the transition 
since $\HH$ jumps and ${1\over \dot\HH-\HH^2}(\HH\Psi +\dot\Psi)$ is 
continuous according to Eqs.~(\ref{mani}) and (\ref{mani2}).

For fluids with adiabatic perturbations and single scalar fields, the 
general solution for the Bardeen potential on super horizon scales is 
given by (see e.g. \cite{mukrep})
\be
 \Psi = A{\HH \over a^2} +B \label{sol}~.
\ee
Here $A$ and $B$ are constants for each mode, depending only on the mode $k$.
During the collapse phase in the ekpyrotic universe, the growing mode 
proportional to $A$ acquires a scale invariant spectrum, $|A|^2k^3=$ const.,
while the spectrum of the constant mode $B$ is blue, 
$|B|^2k^3\propto k^2$ \cite{ekpy3}.
Inserting ansatz~(\ref{sol}) in the continuity condition  (\ref{me1}) for
the metric, yields
\be
B_+\left[\dot\HH_+ -2\HH_+^2\over\dot\HH_+ - \HH_+^2 \right] =
B_-\left[\dot\HH_+ -2\HH_-^2\over\dot\HH_- - \HH_-^2 \right] ~.
\ee
Clearly, since $B_+$ couples only to $B_-$ it inherits the blue 
spectrum of $B_-$. This is the main argument of 
Refs.~\cite{david1,robi,david2}.
However, in this case with $\de f=\rho D_s$, Eq.~(\ref{m3}) leads to 
$\de K\equiv 0$ and does hence not allow for perturbations of the 
surface tension. A possible way out would be to admit anisotropic 
stresses, so that $[\de f/\dot f]_\pm \neq 0$. But we want to clarify the 
attempt taken in Ref~\cite{ekpy3}:

 Another natural coordinate choice is to simply set $f =\eta$ in 
{\em longitudinal gauge}. In this case 
the junction conditions become
\bea
[\Psi]_\pm  &= & 0  \label{mcom1}\\
\left[ \HH\Psi + \dot\Psi \right]_\pm &=& a\ka^2\de p_s \label{mcom2}
\eea
Note that (see e.g.~\cite{du94} Eq. (2.45), 
$4\pi Ga^2\left(\de\rho\right)_{\rm com}= k^2\Psi$ where 
$\left(\de\rho\right)_{\rm com}$ is the energy density perturbation in 
comoving gauge. Hence with this choice the energy density perturbation 
in comoving gauge is continuous through the transition. 

For our general solution (\ref{sol}) this Eqs.~(\ref{mcom1}) 
and (\ref{mcom2}) give
\bean
 A_+/a^2 &=& (A_-/a^2)\HH_-/\HH_+ + (B_- -B_+)/\HH_+ \\
 B_+ &=& {A_-\over a^2}\left[{\HH_+(\dot\HH_- -\HH^2_-) - \HH_-(\dot\HH_+-\HH_+^2) 
	\over 2\HH_+^2 -\dot\HH_+}\right]  \\ &&  +
  B_- \left[ 1 + {\HH_-\HH_+ - \HH_+^2\over 2\HH_+^2-\dot\HH_+}\right]  \\ && +
  \ka^2a\de p_s{\HH_+\over 2\HH_+^2-\dot\HH_+}~.
\eean
 Even if $B_-=\de p_s=0$, $B_+$ will in general not vanish and will 
inherit the flat spectrum of $A_-$. Since the spectra are normalized on 
small scales, the scale invariant contribution from $A_-$ will dominate also if
$B_-\neq 0$ ( unless if its pre-factor $A_-[...]$ vanishes).
This is the matching condition which has been adopted in \cite{ekpy3}.

\section{Conclusions}

We consider the matching conditions which match the energy density
perturbation as 
seen by an observer comoving with the fluid at least as natural as matching 
the energy in longitudinal gauge. Therefore, form the four dimensional point
of view, one cannot decide which one, if any, of these two 
matching conditions is to be preferred. This issue has to be resolved in the
five dimensional picture, which is attempted in Refs.~\cite{ekpy3,nathi}. It 
is also not clear to us how a full, five 
dimensional treatment of perturbation theory would alter this result.
Clearly, since five dimensional perturbations can be re-written as four 
dimensional ones with 'seeds'~\cite{langlois}, the scale invariant mode 
remains as a homogeneous solution. Whether it will generically dominate 
over the 'inhomogeneous modes' is not clear at this stage.
The inhomogeneous modes reflect themselves in this treatment in the 
perturbation of the surface tension $\de p_s$.

The scenario deserves further study.
\vspace{0.3cm}

{\bf Acknowledgment:}\\
I thank Neil Turok, Jean Philippe Uzan and Filippo Vernizzi for discussions.

\end{document}